\documentclass[preprint,showpacs,preprintnumbers,amsmath,amssymb,superscriptaddress,11pt,graphicx]{revtex4}

\usepackage{graphicx}
\usepackage{dcolumn}
\usepackage{bm}
\usepackage{amssymb}

\begin{document}

\title{Relic gravitational waves in the light of 7-year Wilkinson Microwave 
Anisotropy Probe data and improved prospects for the Planck mission}

\author{W.~Zhao}
\email{Wen.Zhao@astro.cf.ac.uk} \affiliation{School of Physics
and Astronomy, Cardiff University, Cardiff, CF24 3AA, United
Kingdom} \affiliation{Wales Institute of Mathematical and
Computational Sciences, Swansea, SA2 8PP, United Kingdom}
\affiliation{Department of
Physics, Zhejiang University of Technology, Hangzhou, 310014,
People's Republic of China}

\author{D.~Baskaran}
\email{Deepak.Baskaran@astro.cf.ac.uk} \affiliation{School of Physics
and Astronomy, Cardiff University, Cardiff, CF24 3AA, United
Kingdom} \affiliation{Wales Institute of Mathematical and
Computational Sciences, Swansea, SA2 8PP, United Kingdom}

\author{L.P.~Grishchuk}
\email{Leonid.Grishchuk@astro.cf.ac.uk} \affiliation{School of Physics
and Astronomy, Cardiff University, Cardiff, CF24 3AA, United
Kingdom} \affiliation{Sternberg Astronomical Institute, Moscow State University, Moscow,
119899, Russia}

\date{\today}


\begin{abstract}

{\small 
The new release of data from Wilkinson Microwave Anisotropy Probe 
improves the observational status of relic gravitational waves. 
The 7-year results enhance the indications of relic gravitational waves in 
the existing data and change to the better the prospects of confident 
detection of relic gravitational waves by the currently operating Planck 
satellite. We apply to WMAP7 data the same methods of analysis that we 
used earlier [W. Zhao, D. Baskaran, and L.P. Grishchuk, Phys. Rev. D {\bf 80}, 
083005 (2009)] with WMAP5 data. We also revised by the same methods our 
previous analysis of WMAP3 data. It follows from the examination of 
consecutive WMAP data releases that the maximum likelihood value of the 
quadrupole ratio $R$, which characterizes the amount of relic gravitational 
waves, increases up to $R=0.264$, and the interval separating this value from 
the point $R=0$ (the hypothesis of no gravitational waves) increases up to a 
$2\sigma$ level. The primordial spectra of density perturbations and 
gravitational waves remain blue in the relevant interval of wavelengths, but 
the spectral indices increase up to $n_s =1.111$ and $n_t=0.111$. Assuming 
that the maximum likelihood estimates of the perturbation parameters that we 
found from WMAP7 data are the true values of the parameters, we find that 
the signal-to-noise ratio $S/N$ for the detection of relic gravitational waves 
by the Planck experiment increases up to $S/N=4.04$, even under pessimistic 
assumptions with regard to residual foreground contamination and instrumental 
noises. We comment on theoretical frameworks that, in the case of success, 
will be accepted or decisively rejected by the Planck observations.}

\end{abstract}



\maketitle

\section{Introduction \label{section1}}

The Wilkinson Microwave Anisotropy Probe (WMAP) Collaboration has 
released the results of 7-year (WMAP7) observations 
\cite{wmap7, wmap7-larson}. In this paper, we apply to WMAP7 data the same 
methods of analysis that we have used before \cite{stable} in the analysis
of WMAP5 data. This is important for updating the present observational 
status of relic gravitational waves and for making more accurate forecasts
for the currently operating Planck mission \cite{planck}. 

In Sec.~\ref{section2.0} we briefly summarize our basic theoretical 
foundations and working tools. Part of this material was present in the 
previous publication \cite{stable}, but in order to make the paper 
self-contained we briefly repeat it here. Section \ref{section2} 
exposes full details of our maximum likelihood analysis of WMAP7 data.
In the focus of attention is the interval of multipoles 
$2 \leq \ell \leq 100$, where gravitational waves compete with density
perturbations. We compare all the results that we derived by exactly the 
same method from WMAP7, WMAP5 and WMAP3 datasets. This comparison demonstrates 
the stability of data and data analysis. On the grounds of this comparison, 
one can say that the perturbation parameters found from the consecutive 
WMAP data releases have the tendency of saturating near some particular 
values. The WMAP7 maximum likelihood (ML) value of the quadrupole ratio $R$ 
is close to previous evaluations of $R$, but increases up to $R=0.264$. The 
interval separating this ML value from the point $R=0$ (the hypothesis 
of no gravitational waves) increases up to a $2\sigma$ level. The primordial 
spectra remain blue, but the spectral indices in the relevant interval of 
wavelengths increase up to $n_s = 1.111$ and $n_t=0.111$.

In Sec.~\ref{section3} we analyze why, to what extent, 
and in what sense our conclusions with respect to relic gravitational 
waves differ from those reached by the WMAP Collaboration. The WMAP team has 
found ``no evidence for tensor modes." A particularly important 
issue, which we discuss in some detail, is the presumed constancy 
(or simple running) of spectral indices. We derive an exact formula for the
spectral index $n_t$ as a function of wavenumbers and discuss in this context 
the formula for running that was used in WMAP analysis. Another contributing 
factor to the difference of conclusions is the difference in our treatments of 
the inflationary ``consistency relations" based on the inflationary ``classic 
result." We do not use the inflationary theory.

A comprehensive forecast for Planck findings in the area of relic 
gravitational waves is presented in Sec.~\ref{section4}. We discuss
the efficiency of various information channels, i.e. various
correlation functions and their combinations. We perform multipole 
decomposition of the calculated $S/N$ and discuss physical implications of 
the detection in various intervals of multipole moments. We stress again 
that the $B$-mode detection provides the most of $S/N$ only in the conditions 
of very deep cleaning of foregrounds and relatively small values of $R$. 
The improvements arising from a 28-month, instead of a nominal 14-month, 
Planck survey are also discussed. In the center of our attention is the
model with the WMAP7 maximum likelihood set of parameters. For this model, 
the signal-to-noise ratio $S/N$ in the detection of relic gravitational waves 
by Planck experiment increases up to $S/N=4.04$, even under pessimistic 
assumptions with regard to residual foreground contamination and 
instrumental noises. Section \ref{bayescomp} gives Bayesian 
comparison of different theoretical frameworks and identifies predictions  
of $R$ that may be decisively rejected by the Planck observations.

\section{Perturbation parameters and CMB power spectra \label{section2.0}}

The temperature and polarization anisotropies of CMB are produced by density 
perturbations, rotational perturbations and gravitational waves. Rotational
perturbations are expected to be very small and are usually ignored, and are in 
this paper, too. The cosmological perturbations are characterized by their 
gravitational field (metric) power spectra which are in general functions of 
time. Here, we introduce the notations and equations that will be used in 
subsequent calculations.

As before (see \cite{stable}, \cite{mg12}, and references therein), we are 
working with perturbed Friedmann-Lemaitre-Robertson-Walker universes
\begin{eqnarray}
ds^2=-c^2dt^2+a^2(t)(\delta_{ij}+h_{ij})dx^idx^j 
=a^2(\eta)[-d\eta^2+(\delta_{ij}+h_{ij})dx^idx^j],
\nonumber
\label{metric}
\end{eqnarray}
where the functions $h_{ij}(\eta,{\bf x})$ are metric perturbation 
fields. Their spatial Fourier expansions are given by
\begin{eqnarray}
\label{h}
 h_{ij}(\eta,{\bf x})=\frac{\mathcal{C}}{(2\pi)^{3/2}}
\int_{-\infty}^{+\infty}\frac{d^3{\bf
 n}}{\sqrt{2n}} \sum_{s=1,2}\left[\stackrel{s}{p}_{ij}({\bf n})
 \stackrel{s}{h}_{n}(\eta)e^{ i{\bf n}\cdot{\bf x}}
\stackrel{s}{c}_{\bf n}+ \stackrel{s~*}{p_{ij}}({\bf n})
\stackrel{s~*}{h_{n}}(\eta)e^{-i{\bf n}\cdot{\bf x}}
\stackrel{s~\dag}{c_{\bf n}} \right].
\end{eqnarray}
The polarization tensors $\stackrel{s}{p}_{ij}({\bf n})$ ($s=1,2$) refer 
either to the two transverse-traceless components of gravitational waves 
(gw) or to the scalar and longitudinal-longitudinal components of 
density perturbations (dp). Density perturbations necessarily include 
perturbations of the accompanying matter fields (not shown here).

In the quantum version of the theory, the quantities  
$\stackrel{s}{c}_{\bf n}$ and $\stackrel{s~\dag}{c_{\bf n}}$ are the annihilation 
and creation operators, respectively, of the considered type of perturbations, and 
the $|0 \rangle$ is the initial vacuum (ground) state of the corresponding 
time-dependent Hamiltonian. The metric power spectrum $h^2(n,\eta)$ is 
defined by the expectation value of the quadratic combination of the 
metric field:
\begin{eqnarray}
\langle 0| h_{ij}(\eta,{\bf x})h^{ij}(\eta,{\bf x}) |0 \rangle
= \int\limits_{0}^{\infty} \frac{dn}{n}~h^2(n,\eta),~~~~~
\label{powerspectrumhdef}
h^2(n,\eta) \equiv \frac{\mathcal{C}^2}{2\pi^2}n^2
\sum_{s=1,2}|\stackrel{s}{h}_n(\eta)|^2.
\end{eqnarray} 
The mode functions $\stackrel{s}{h}_{n}(\eta)$ are taken either from gw 
or dp equations, and $\mathcal{C}=\sqrt{16\pi}l_{\rm Pl}$ for gravitational
waves and $\mathcal{C}=\sqrt{24\pi}l_{\rm Pl}$ for density perturbations.

The simplest assumption about the initial stage of cosmological expansion (i.e. 
about the initial kick that presumably took place soon after the birth of our 
Universe \cite{zg,gr09}) is that it can be described by a single power-law 
scale factor \cite{gr74,disca,gr09}
\begin{equation}
\label{inscf}
a(\eta) = l_o|\eta|^{1+ \beta},
\end{equation}
where $l_o$ and $\beta$ are constants, $\beta < -1$. Then the generated 
primordial power spectra (primordial means the interval of the spectrum 
pertaining to wavelengths longer than the Hubble radius at the considered 
moment of time) have the universal power-law dependence, both for gw and dp:
\begin{equation}
\nonumber
h^{2}(n) \propto n^{2(\beta+2)}.
\end{equation}
It is common to write these power spectra separately for gw and dp: 
\begin{equation}
\label{primsp}
h^2(n)~({\rm gw}) = B_t^2 n^{n_t}, ~~~~~h^2(n)~({\rm dp})=B_s^2 n^{n_s -1}.
\end{equation}
In accordance with the theory of quantum-mechanical generation of cosmological
perturbations \cite{gr74,disca,gr09}, the spectral indices are 
approximately equal $n_s-1 = n_t = 2(\beta+2)$ and the amplitudes $B_t$ and
$B_s$ are of the order of magnitude of the ratio $H_i/H_{\rm Pl}$, where 
$H_i\sim c/l_o$ is the characteristic value of the Hubble parameter during the 
kick.

If the initial stage of expansion is not assumed to be a pure power-law 
evolution (\ref{inscf}), the spectral indices $n_t$ and $n_s-1$ are not 
constants, but their wavenumber dependence is calculable from the 
time dependence of the scale factor $a(\eta)$ and its time derivatives 
\cite{grsol}. In fact, as we shall argue below, the CMB 
data suggest that even at a span of 2 orders of magnitude in terms of 
wavelengths the spectral index $n_s$ is not the same. We discuss this issue 
in detail in Sec.~\ref{section3}.

In what follows, we use the numerical code CAMB \cite{CAMB} 
and related notations for gw and dp power spectra adopted there:
\begin{eqnarray}
\label{PsPt}
P_{t}(k)=A_t \left(\frac{k}{k_0}\right)^{n_t},~~~
P_{s}(k)=A_s \left(\frac{k}{k_0}\right)^{n_s-1},
\end{eqnarray}
where $k_0=0.002$Mpc$^{-1}$. Technically, the power spectrum $P_{s}(k)$ refers
to the curvature perturbation called $\mathcal{R}$ or $\zeta$, but the 
amplitudes $B_s$ and $(A_s)^{1/2}$ are equal to each other up to a numerical 
coefficient of order 1. The constant dimensionless wavenumber $n$ is related 
to the dimensionful $k$ by $k= n/(2 l_{H})$, where $l_{H}=c/H_0$ is the 
present-day Hubble radius. The wavenumber $n=n_H=4\pi$ marks the wavelength 
equal to $l_H$ today. The CMB temperature anisotropy at the multipole $\ell$ 
is mostly generated by metric perturbations with wavenumbers
$n\approx\ell$ (see \cite{zbg} for details). Setting $h=0.704$ we obtain 
$\ell\approx(0.85\times 10^{4}$Mpc$)k$, which is consistent with
the numerical result $\ell\approx (1.0\times 10^{4}$Mpc$)k$ derived in 
\cite{zb}.

The CMB temperature and polarization anisotropies are usually characterized 
by the four angular power spectra: $C_{\ell}^{TT}$, $C_{\ell}^{EE}$, 
$C_{\ell}^{BB}$, and $C_{\ell}^{TE}$ as functions of the multipole $\ell$. The 
contribution of gravitational waves to these power spectra has been studied, 
both analytically and numerically, in a number of papers 
\cite{Polnarev1,a8,a11,a12,a13}. The derivation of today's CMB 
power spectra brings us to approximate formulas of the following structure 
\cite{a12}:
\begin{eqnarray}
\label{exact-clxx'} 
\begin{array}{l} 
C_{\ell}^{TT}= \int \frac{dn}{n} h^2(n, \eta_{rec}) 
\left[F^T_{\ell}(n)\right]^2, \\
C_{\ell}^{TE}=\int \frac{dn}{n} h(n, \eta_{rec}) h^{\prime}(n, \eta_{rec}) 
\left[F^T_{\ell}(n) F^E_{\ell}(n)\right], \\
C_{\ell}^{YY}=\int \frac{dn}{n} (h^{\prime})^2 (n, \eta_{rec})
\left[F^Y_{\ell}(n)\right]^2, ~~~~{\rm where}~Y=E,B.
\end{array}
\end{eqnarray}
In the above expressions, $h^2(n, \eta_{rec})$ and 
$(h^{\prime})^2 (n, \eta_{rec}$) are the power spectra of the gravitational 
wave field and its first time-derivative, respectively. The spectra are taken 
at the recombination (decoupling) time 
$\eta_{rec}$. The functions $F^X_{\ell}(n)$ ($X=T, E, B$) take care of the 
radiative transfer of CMB photons in the presence of metric perturbations. As 
already mentioned, the power residing in the metric fluctuations at wavenumber 
$n$ mostly translates into the CMB $TT$ power at the angular scales 
$\ell \approx n$. Similar results hold for the CMB power spectra induced by 
density perturbations. The actually performed numerical calculations use 
equations more accurate than Eq.~(\ref{exact-clxx'}). They also include the
effects of the reionization era.

The CMB power spectra needed for the analysis of WMAP data are 
calculated in the framework of background cosmological $\Lambda$CDM model 
characterized by the WMAP7 best-fit parameters \cite{wmap7} 
\begin{eqnarray}
 \label{background}
 \begin{array}{c}
 \Omega_bh^2=0.02260,~~~\Omega_ch^2=0.1123, ~~~
 \Omega_{\Lambda}=0.728,~~~\tau_{\rm reion}=0.087,~~~h=0.704.
 \end{array}
\end{eqnarray}

To quantify the contribution of relic gravitational waves to the CMB we use 
the quadrupole ratio $R$ defined by
\begin{eqnarray}
\label{defineR}
R \equiv \frac{C_{\ell=2}^{TT}({\rm gw})}{C_{\ell=2}^{TT}({\rm dp})}.
\end{eqnarray}
Another measure is the so-called tensor-to-scalar ratio $r$. This quantity is 
constructed from the primordial power spectra (\ref{PsPt})
\begin{eqnarray}
\label{definer}
 r \equiv \frac{A_t}{A_s}.
\end{eqnarray}
Often this parameter is linked to incorrect (inflationary) statements, such as
the inflationary `consistency relation' (more details in Sec.~\ref{section3}). 
However, if one uses $r$ without implying inflationary claims, one can find a 
useful relation between $R$ and $r$.

\begin{figure}
\begin{center}
\includegraphics[height=10cm]{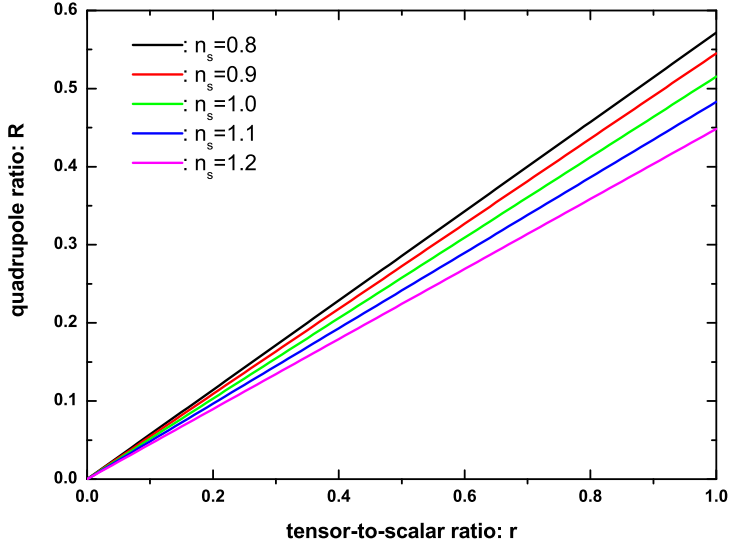}
\end{center}
\caption{The relation between $R$ and $r$ for different values of the spectral 
index $n_s$. From top to bottom the $n_s$ changes from $n_s=0.8$ to $n_s=1.2$.} 
\label{fig0}
\end{figure}

In general, the relation between $R$ and $r$ depends on background cosmological 
parameters and spectral indices $n_s$ and $n_t$. We found this relation 
numerically using the CAMB code \cite{CAMB}. (For a semianalytical approach 
see \cite{rRrelation}.) The results are plotted in Fig.~\ref{fig0}.
For this calculation we used the background cosmological
parameters (\ref{background}) and the condition $n_t=n_s-1$ required by the 
theory of quantum-mechanical generation of cosmological perturbations. 
We verified that the relation $r(R)$ only weakly depends on the background
parameters and does not change significantly when the values 
(\ref{background}) are varied within the WMAP7 $1\sigma$ error range
\cite{wmap7}. It is seen from the graph that $r = 1.92 R$ in the case of 
$n_s=1.0$, and $r\approx 2R$ for all considered $n_s$, if $R$ and $r$ are 
sufficiently small. In other words, one can use $r\approx 2R$ for a quite 
wide class of models.

We do not know enough about the very early Universe to predict $R$ with any 
certainty. However, since the theory of quantum-mechanical generation of 
cosmological perturbations requires that the amplitudes $B_t$ and $B_s$ (as 
well as $A_t$ and $A_s$ in Eq.(\ref{PsPt})) should 
be of the same order of magnitude, our educated guess is that $R$ should lie 
somewhere in the range $R\in[0.01,1]$. If $R$ were observationally found 
significantly outside this range, we would have to conclude that the underlying 
perturbations are unlikely to be of quantum-mechanical origin. On the other 
hand, the most advanced inflationary theories predict the ridiculously small 
amounts of gravitational waves, something at the level of $r \approx 10^{-24}$ 
or less, $r\in[0,10^{-24}]$. The rapidly improving CMB data will soon allow 
one to decisively discriminate between these theoretical frameworks (let alone 
the already performed discrimination on the grounds of purely theoretical 
consistency). We discuss this issue in Sec.~\ref{bayescomp}.

\section{Evaluation of relic gravitational waves from 7-year WMAP data 
\label{section2}}

\subsection{Likelihood function \label{section2.1}}

Relic gravitational waves compete with density perturbations in
generating CMB temperature and polarization anisotropies at
relatively low multipoles. For this reason we focus on the WMAP7 
data at $2 \le \ell\le100$. As before \cite{zbg,stable}, 
the quantities $D_{\ell}^{TT}$, $D_{\ell}^{TE}$, $D_{\ell}^{EE}$,
and $D_{\ell}^{BB}$ denote the estimators (and also the actual observed data 
points in the likelihood analysis) of the corresponding power spectra. 
Since the WMAP7 $EE$ and $BB$ 
observations are not particularly informative, we marginalize (integrate) 
the total probability density function (pdf) over the variables 
$D_{\ell}^{EE}$ and $D_{\ell}^{BB}$ \cite{zbg,stable}. The resulting pdf 
\cite{stable} is a function of $D_{\ell}^{TT}$ and $D_{\ell}^{TE}$: 
\begin{eqnarray}
f(D_{\ell}^{TT},D_{\ell}^{TE})= n^2{x}^{\frac{n-3}{2}}
\left\{2^{1+n}\pi\Gamma^2(\frac{n}{2})(1-\rho_{\ell}^2)(\sigma_{\ell}^T)^{2n}
(\sigma_{\ell}^E)^2\right\}^{-\frac{1}{2}}
\nonumber\\
 \times\exp\left\{\frac{1}{1-\rho^2_{\ell}}\left(\frac{{\rho_{\ell}}
{z}}{{\sigma_\ell^T}{\sigma_\ell^E}}-\frac{{z}^2}{2x{(\sigma_\ell^E)^2}
}-\frac{{x}}{2{(\sigma_\ell^T)^2}}\right)\right\}.
 \label{pdf_CT}
\end{eqnarray}

This pdf contains the variables $D_{\ell}^{XX'}$ ($XX'=TT,TE$) through the 
quantities ${x}\equiv n(D_\ell^{TT}+N_{\ell}^{TT})$ and 
${z}\equiv nD_\ell^{TE}$, where 
$N_{\ell}^{TT}$ and $N_{\ell}^{EE}$ are total noise power spectra. 
Information about the power spectra $C_{\ell}^{XX'}$ is contained in the
quantities ${\sigma_\ell^T}$, ${\sigma_\ell^E}$ and $\rho_{\ell}$
(see \cite{stable} for details). The sought after parameters $A_s$, $A_t$, 
$n_s$, and $n_t$ enter the pdf through the $C_{\ell}^{XX'}$. The quantity 
$n=(2\ell+1)f_{\rm sky}$ in (\ref{pdf_CT}) is the effective number of degrees 
of freedom at multipole $\ell$, where $f_{\rm sky}$ is the sky-cut factor.

In the WMAP7 data release the sky-cut factor is $f_{\rm sky}=0.783$ 
\cite{wmap7-larson}, which is slightly smaller than $f_{\rm sky}=0.85$ used 
in WMAP5 data analysis \cite{stable}. The smaller $f_{\rm sky}$ increases the 
uncertainties, but this disadvantage is more than compensated by the reduction 
of overall noises. Therefore, the error bars surrounding the WMAP7 data points 
are somewhat smaller than those for the WMAP5 data release. This fact,
together with slightly shifted data points themselves, allows us to 
strengthen our conclusions (see below) about the presence of gravitational 
wave signal in the WMAP data.

We seek the perturbation parameters $R$, $A_s$ and $n_s$ ($n_t=n_s-1$) along 
the lines of our maximum likelihood analysis of WMAP5 data \cite{stable}.
The pdf (\ref{pdf_CT}) considered as a function of unknown $R$, $A_s$, and $n_s$ 
with known data points $D_{\ell}^{XX'}$ is a likelihood function subject to 
maximization. For a set of observed multipoles 
$\ell=2,\cdot\cdot\cdot,\ell_{max}$, the likelihood function can be 
written as \cite{stable}
\begin{eqnarray}
 \label{ctlikelihood2}
 -2\ln \mathcal{L}=\sum_{\ell}\left\{\frac{1}{1-\rho^2_{\ell}}
\left(\frac{{z}^2}{x{(\sigma_\ell^E)^2} }+
\frac{{x}}{{(\sigma_\ell^T)^2}} -\frac{{2\rho_{\ell}}{z}}
{{\sigma_\ell^T}{\sigma_\ell^E}}\right) +
\ln\left((1-\rho_{\ell}^{2})(\sigma_{\ell}^T)^{2n}(\sigma_{\ell}^E)^2\right)
\right\}+C,
\end{eqnarray} 
where the constant $C$ is chosen to make the maximum value of $\mathcal{L}$ 
equal to 1.

\subsection{Results of the analysis of the WMAP7 data\label{results}}

\begin{figure}
\begin{center}
\includegraphics[width=6cm,height=7cm]{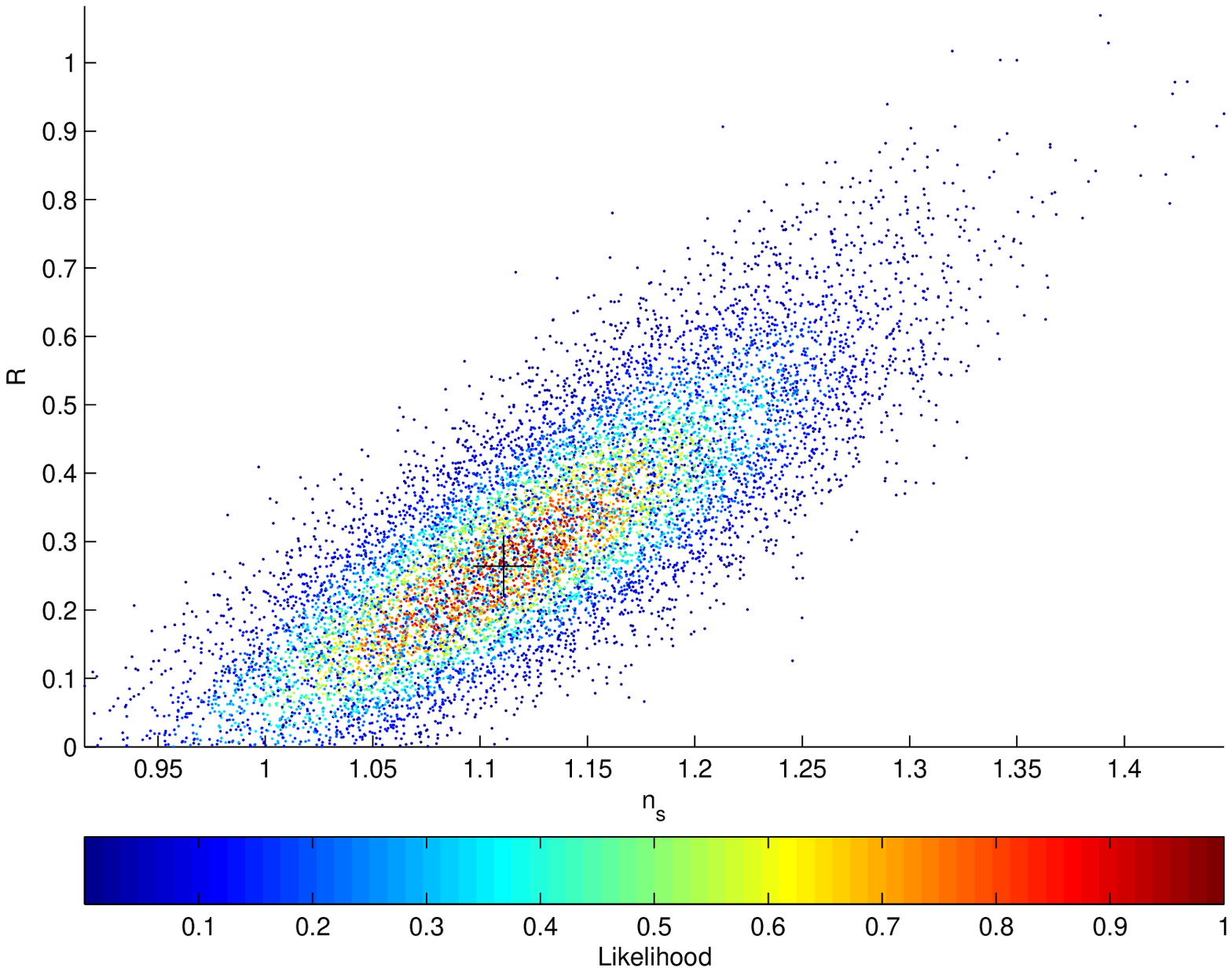}
\includegraphics[width=6cm,height=7cm]{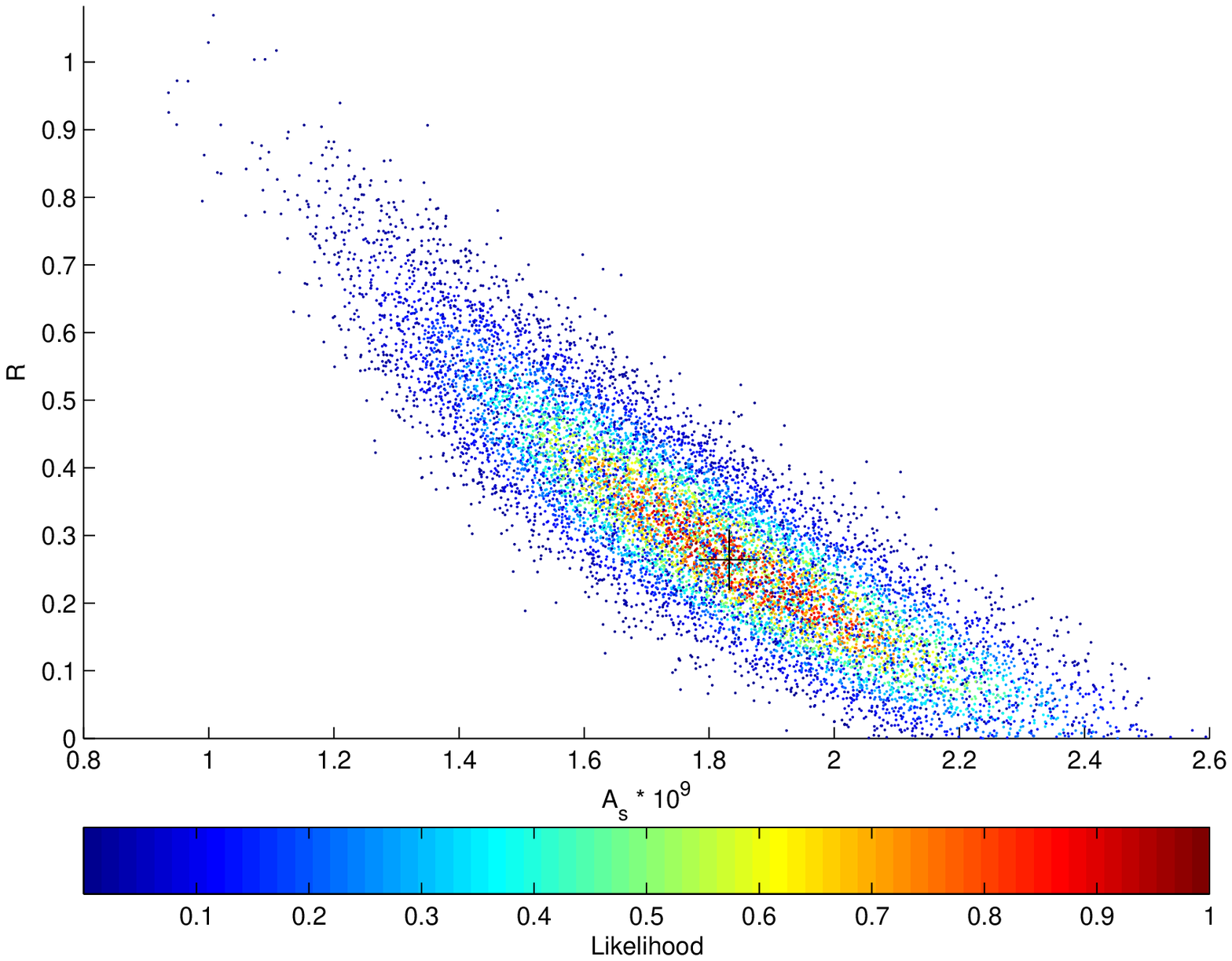}
\end{center}
\caption{The projection of 10,000 samples of the 3-dimensional
likelihood function, based on the WMAP7 data, onto the  $R-n_s$
(left panel) and $R-A_s$ (right panel) planes. The black $+$
indicates the maximum likelihood parameters listed in (\ref{best-fit}). }
\label{figurea1.1}
\end{figure}

The WMAP7 data points for $D_{\ell}^{TT}$ and $D_{\ell}^{TE}$ at multipoles 
$2\le \ell \le\ell_{max}=100$ were taken, with gratitude, from the Website
\cite{LAMBDA}. The 3-dimensional likelihood function (\ref{ctlikelihood2}) was 
probed by the Markov chain Monte Carlo method \cite{mcmc1, mcmc2} using 
10,000 samples. The ML values of the three perturbation 
parameters $R$, $n_s$ and $A_s$ were found to be
\begin{eqnarray}
R=0.264,~~~n_s=1.111,~~~A_s=1.832\times10^{-9} 
\label{best-fit}
\end{eqnarray}
and $n_t=0.111$. In Fig.~\ref{figurea1.1} we show the projection of the 
10,000 sample points on the 2-dimensional planes $R-n_s$ and $R-A_s$. The 
color of an individual point signifies the value of the 3-dimensional 
likelihood of the corresponding sample. The projections of the maximum 
(\ref{best-fit}) are shown by a black $+$. (The value $R=0.264$ is equivalent
to $r=0.550$.)

\begin{figure}
\begin{center}
\includegraphics[width=6cm,height=7cm]{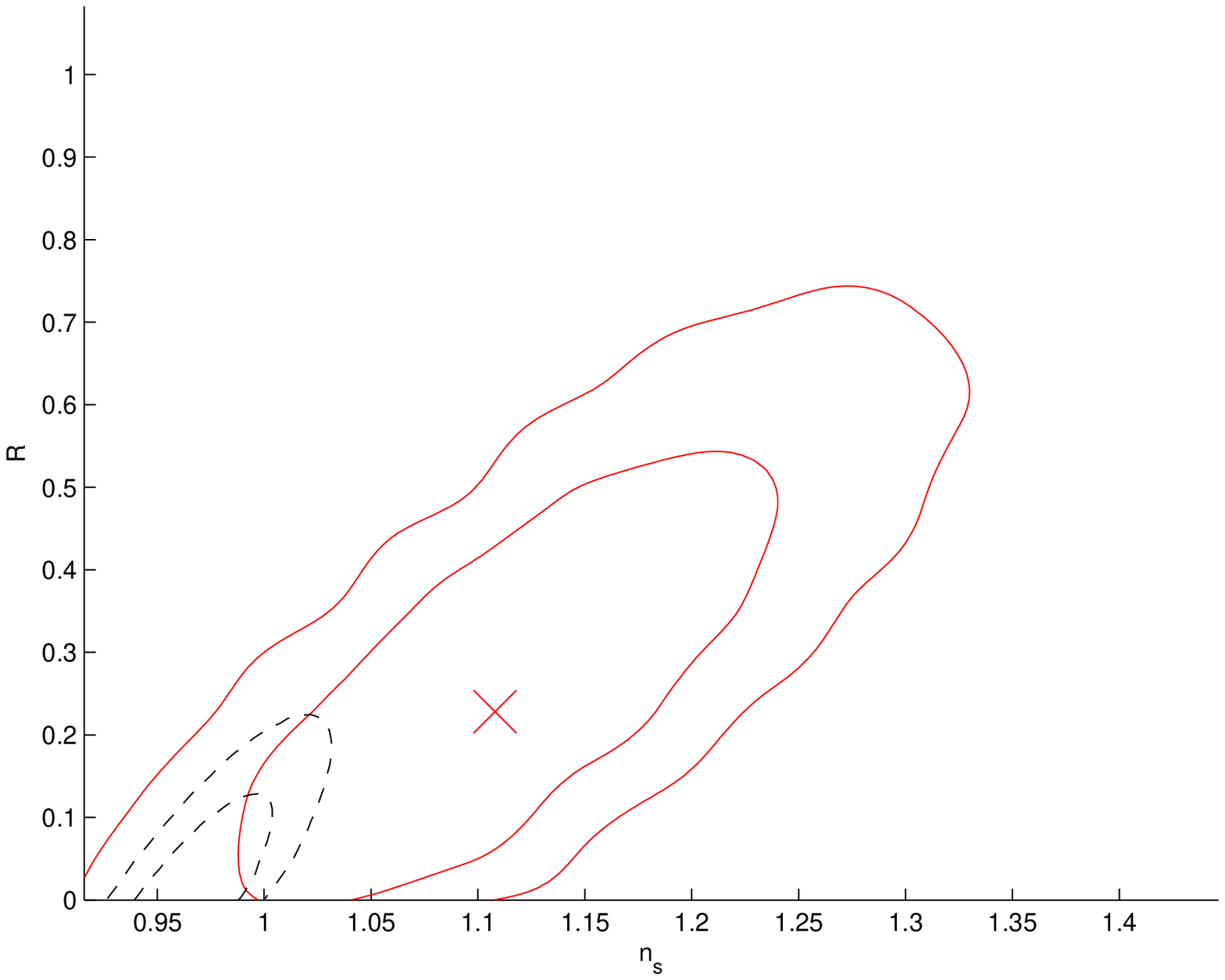}
\includegraphics[width=6cm,height=7cm]{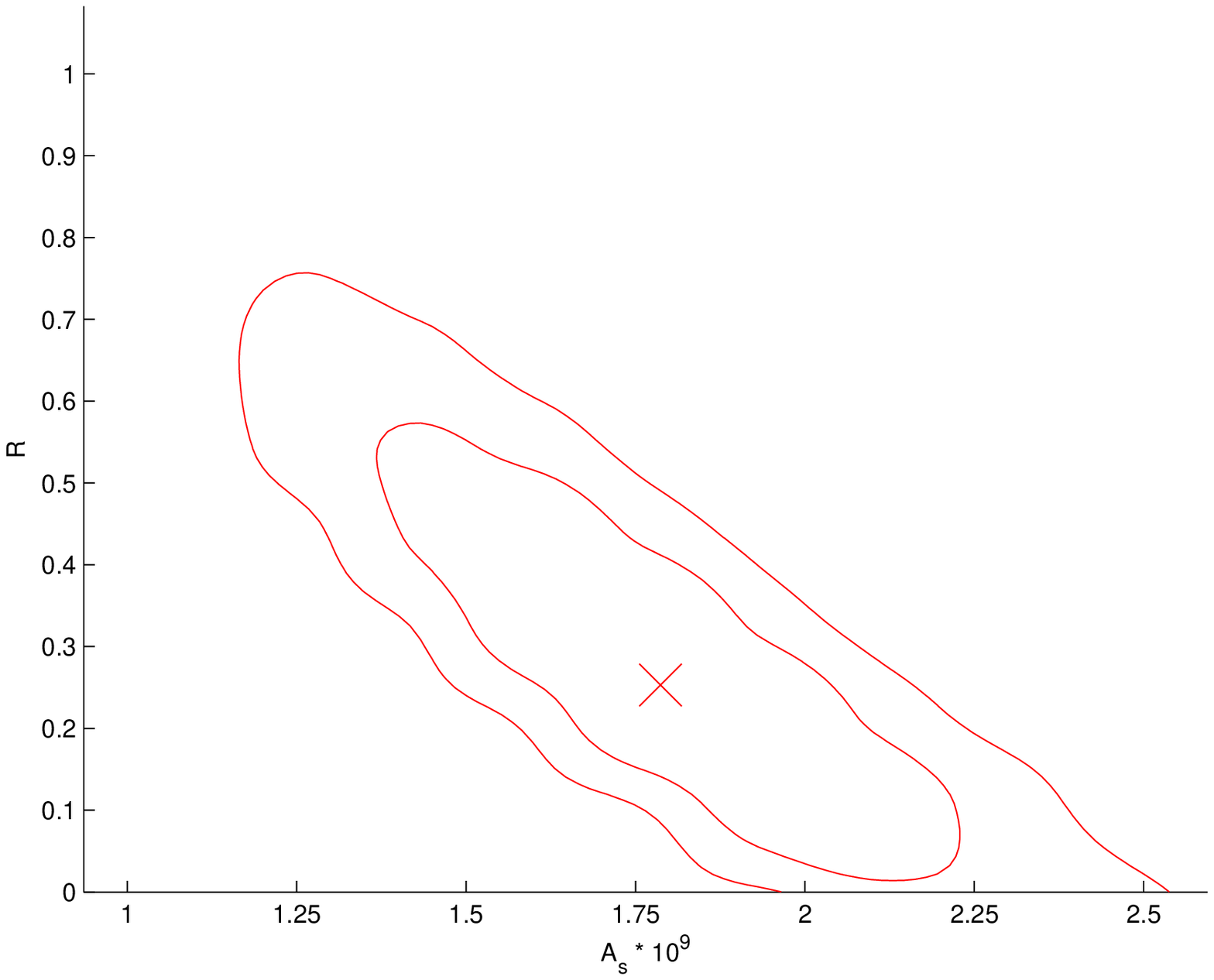}
\end{center}
\caption{The ML point (red $\times$) and the $68.3\%$ and $95.4\%$
confidence intervals (red solid lines) for the
2-dimensional distributions $R-n_s$ (left panel) and $R-A_s$ (right panel).
The left panel shows also the WMAP7 confidence contours (black dashed line) 
derived under the assumption that the spectral index $n_s$ is one and the same
constant throughout all measured multipoles \cite{wmap7,wmap7-larson}. The WMAP7 papers do
not show the confidence contours in the $R-A_s$ plane.} \label{fig2d}
\end{figure}

Before analyzing the 3-parameter results, it is instructive to consider 
the 2-parameter and 1-parameter probability distributions of the sought 
after parameters. These marginalized distributions are obtained by integrating 
the likelihood function $\mathcal{L}$ (\ref{ctlikelihood2}) (already represented 
by 10,000 points) over one or two parameters. By integrating over $A_s$ or 
$n_s$, we arrive at 2-dimensional distributions for the pairs $R-n_s$ or 
$R-A_s$, respectively. The area around the maximum of the resulting 
distributions is shown in Fig.~\ref{fig2d}. In the $R-n_s$ space, the maximum 
is located at 
\begin{eqnarray} 
\label{2dRn}
R=0.228,~~~n_s=1.108.
\end{eqnarray}
In the $R-A_s$ space, the maximum is located at
\begin{eqnarray} 
\label{2dRA}
R=0.253,~~~A_s=1.787\times10^{-9}.
\end{eqnarray}
In the left panel we also reproduce the 2-dimensional contours obtained by the 
WMAP team \cite{wmap7,wmap7-larson} (their $r$ is translated into our $R$). In contrast to 
the WMAP5 paper \cite{wmap5}, the WMAP7 paper \cite{wmap7-larson} shows only 
the uncertainty contours derived under the assumption of a strictly constant 
spectral index $n_s$, but not for the case of its running.

\begin{figure}
{\includegraphics[height=10cm]{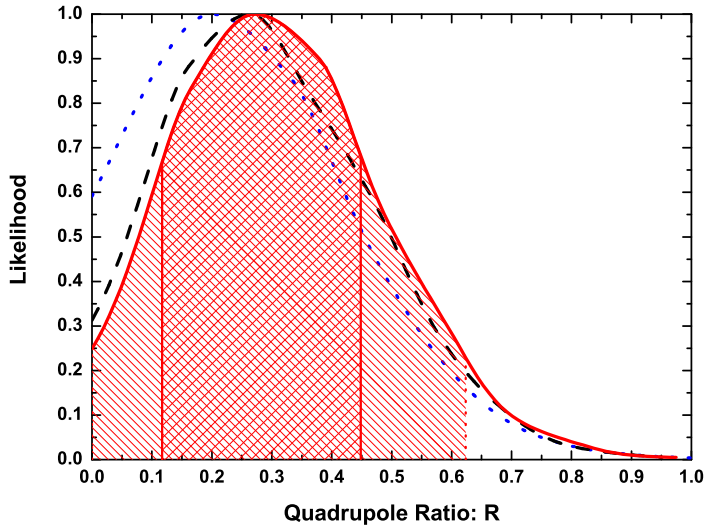}} \caption{The 1-dimensional
likelihoods for $R$. The results of the analysis of WMAP7, WMAP5 and WMAP3 
data are shown, respectively, by the red (solid), black (dashed), and blue 
(dotted) curves. The shaded regions indicate the $68.3\%$ and $95.4\%$ 
confidence intervals for WMAP7 likelihood.}\label{figureb12}
\end{figure}

\begin{figure}
{\includegraphics[width=18cm,height=10cm]{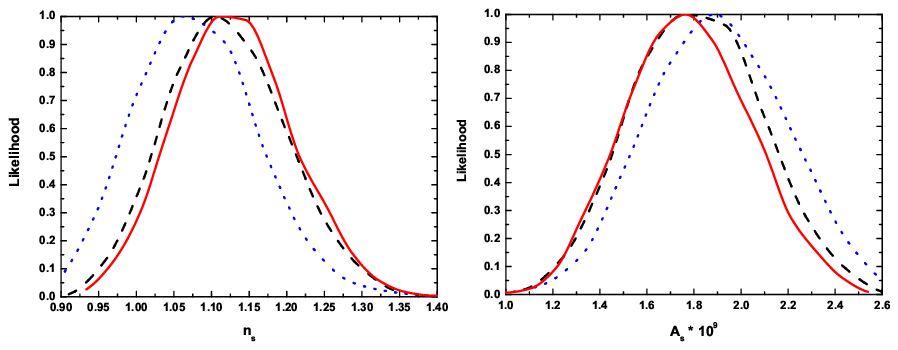}} \caption{The 1-dimensional
likelihoods for $n_s$ (left) and $A_s$ (right). In both panels the
red (solid), black (dashed) and blue (dotted) curves show 
the results for WMAP7, WMAP5 and WMAP3 data, respectively.}
\label{figureb13}
\end{figure}

The 1-dimensional distributions for  $R$, $n_s$ or $A_s$ are obtained by 
integrating the likelihood function $\mathcal{L}$ (\ref{ctlikelihood2})
over the sets of two parameters ($A_s$, $n_s$), ($A_s$,
$R$) or ($R$, $n_s$), respectively. These distributions are presented in
Figs.~\ref{figureb12} and ~\ref{figureb13} (red solid lines). The
ML values of the parameters and their $68.3\%$ confidence intervals are 
found to be
\begin{eqnarray}
\label{best-fit-1d}
 R=0.273^{+0.185}_{-0.156},~~n_s=1.112^{+0.089}_{-0.064} ,
~~A_s=(1.765^{+0.279}_{-0.263})\times10^{-9}.
\end{eqnarray}
For completeness and comparison (see Sec.~\ref{compar}), we also show in 
Figs.~\ref{figureb12} and ~\ref{figureb13} the 1-dimensional (1-D, for 
brevity) distributions derived by exactly the same procedure from WMAP5 and 
WMAP3 observations. Obviously, the WMAP5 curves are copies of the previously 
reported distributions \cite{stable}; the WMAP3 curves are explained in 
Sec.~\ref{compar}.

The constancy of spectral indices is a matter of special discussion in
Sec.~\ref{section3}. In preparation for this discussion, we report the results
of our likelihood analysis of data in other intervals of multipole moments 
$\ell$, in addition to the interval $2\le\ell\le100$ that resulted in the
ML values (\ref{best-fit}). First, we analyzed the WMAP7 data in the interval 
$101\leq\ell\leq220$. The $n_s$ coordinate of the maximum in 
3-dimensional space $R, A_s$, and $n_s$ was found to be $n_s=0.951$. The 1-D 
marginalized distribution for $n_s$ gave the ML result 
\begin{eqnarray}
\label{1ns101}
n_s=0.969^{+0.083}_{-0.063} ~(68.3\%~ {\rm C.L.}). 
\end{eqnarray}
Second, we have done the same in the combined interval of multipoles 
$2\le\ell\leq220$. The maximum of 3-D likelihood lies at 
$n_s=1.003$, whereas the 1-D ML result is 
\begin{eqnarray}
\label{1ns220}
n_s=1.021^{+0.043}_{-0.038} ~ (68.3\% ~{\rm C.L.}). 
\end{eqnarray}
The 1-D results allow one to make easier comparison of confidence intervals. 
As is seen from (\ref{best-fit-1d}) and (\ref{1ns101}), the 1-D determinations
of $n_s$ in the adjacent ranges $2\le\ell\le100$ and $101\leq\ell\leq220$ 
overlap, but only marginally, in their $1\sigma$ intervals. As
expected, if $n_s$ is assumed constant in the entire range $2\le\ell\le220$, 
its value (\ref{1ns220}) is intermediate between (\ref{best-fit-1d}) and 
(\ref{1ns101}). The spreads of wavenumbers to which the 3-d spectral indices 
$n_s=1.111$, $n_s=0.951$ and $n_s=1.003$ refer are shown by marked red lines 
in Fig.~\ref{sec2-fig2}.

\subsection{Comparison of results derived from WMAP7, WMAP5 and WMAP3 data
releases\label{compar}}

One and the same 3-parameter analysis of WMAP5 and WMAP7 data has resulted in 
somewhat different ML parameters. From WMAP7 observations we extracted the 
ML parameters (\ref{best-fit}), whereas the ML parameters extracted from 
WMAP5 observations \cite{stable} are 
\begin{eqnarray}
\nonumber
R=0.229,~~~n_s=1.086,~~~A_s=1.920\times10^{-9}.
\end{eqnarray}
Certainly, the results are 
consistent and close to each other. The same holds true for marginalized 
1-dimensional parameters and distributions shown in Figs.~\ref{figureb12} 
and ~\ref{figureb13}. This similarity of results testifies to the stability 
of data and data analysis. There exist, however, important trends in the 
sequence of ML parameters extracted from the progressively improving WMAP3, 
WMAP5, and WMAP7 data. We want to discuss these trends.

Specially for this discussion, we derived the parameters $R, n_s$, and $A_s$ 
from WMAP3 data in exactly the same manner as was done here 
and in \cite{stable} with WMAP7 and WMAP5 data releases. 
Previously \cite{zbg}, we derived these parameters by a different method:
we restricted the likelihood analysis to $TE$ data and a single parameter 
$R$, while $n_s$ and $A_s$ were determined from phenomenological 
relations designed to fit the $TT$ data. That analysis has led us 
to $R=0.149^{+0.247}_{-0.149}$ and $n_s=1.002$. Our new 
derivation, based on 3-dimensional likelihood, gives the following 
WMAP3 maximum likelihood parameters: 
\begin{eqnarray}
\nonumber
R=0.181,~~~n_s=1.045,~~~A_s=2.021\times10^{-9}.
\end{eqnarray}
The corresponding 1-dimensional distributions give
\begin{eqnarray}
\label{1dwmap3}
R=0.205^{+0.181}_{-0.157}, ~~n_s=1.059^{+0.097}_{-0.066},
~~A_s=(1.894^{+0.290}_{-0.307})\times10^{-9}.
\end{eqnarray}
The 1-dimensional WMAP3 distributions are plotted in Figs. \ref{figureb12} 
and ~\ref{figureb13} by dotted curves alongside with WMAP5 (dashed) and
WMAP7 (solid) curves.

Looking at all derivations and graphs collectively, we can draw the 
following conclusions. First, about $R$. The maximum likelihood value of $R$ 
increases as one goes over from 3-year to better quality 7-year data. The ML 
value $R=0.264$ of WMAP7 data release is $15\%$ larger than the analogous 
result $R=0.229$ obtained from WMAP5 data. The 3-, 2-, and 1-parameter 
determinations of $R$ persistently concentrate somewhere around the mark 
$R=0.25$. The uncertainties $\Delta R$, although still considerable, get 
smaller as one progresses from 3-year to better quality 7-year data. The $R=0$ 
(no gravitational waves) hypothesis is under increasing pressure. For example, 
the WMAP7 1-parameter result, Eq.(\ref{best-fit-1d}) and Fig.~\ref{figureb12}, 
excludes the $R=0$ hypothesis at almost $2\sigma$ level. To be more precise, 
the $R=0$ point is right on the boundary of the $94\%$ confidence area 
($94\%$ of surface area under the WMAP7 curve in Fig.~\ref{figureb12})
surrounding the 1-D ML value $R=0.273$. This is an improvement in 
comparison with a slightly larger than $1\sigma$ interval separating $R=0$
from the WMAP5 1-d ML value $R=0.266$  \cite{stable}. 
Admittedly, the gradual changes in ML values of $R$ are small, while 
uncertainties are still large. It would not be very surprising if random 
realizations of noise moved the consecutive 3-year, 5-year, 7-year ML values of $R$ 
in arbitrary order. Nevertheless, we observe the tendency for systematic 
increase and saturation of $R$ alongside with decrease of $\Delta R$.

Second, our conclusions about $n_s$. Together with the increase of ML $R$, we 
observe the tendency for the increase of $n_s$ accompanied by the 
theoretically expected and understood decrease of $A_s$ (the contribution of 
relic gravitational waves becomes larger, while the contribution of density 
perturbations becomes smaller). The ML value $n_s=1.111$ derived from the
WMAP7 data, Eq.(\ref{best-fit}), is larger than $n_s=1.086$ derived from 
WMAP5 data, while the WMAP7 value of $A_s$ is somewhat smaller than that of 
WMAP5 \cite{stable}. The tendency for increase of $n_s$ and decrease of
$A_s$ is also illustrated by the 1-parameter distributions in 
Fig.~\ref{figureb13}. The spectral indices $n_s, n_t$ persistently point out 
to blue primordial spectra, i.e. $n_s >1$ for density perturbations and 
$n_t>0$ for gravitational waves, in the interval of wavelengths responsible 
for CMB anisotropies at $2\le\ell\le100$. The larger values of $n_s,~n_t$ 
derived from WMAP7 data release enhance the doubt on whether the conventional 
scalar fields could be the driver of the initial kick, since these cannot 
support $\beta >-2$ in Eq.(\ref{inscf}) and, consequently, $n_s>1,~ n_t >0$
in Eq.(\ref{primsp}). (Since the inflationary theory is capable of predicting
virtually anything that one can possibly imagine, there exist of course
literature claiming that inflation can predict blue spectra. But these 
claims are based on the incorrect (inflationary) formula for density 
perturbations, see \cite{disca}.) The questions pertaining to the spectral 
indices are analyzed in some detail in the next section.

\section{Comparison of our results with conclusions of WMAP collaboration
\label{section3}}

Having analyzed the 7-year data release, the WMAP collaboration concludes that 
a minimal cosmological model without gravitational waves and with a constant
spectral index $n_s$ across the entire interval of considered wavelengths 
remains a ``remarkably good fit" to ever improving CMB data and other datasets. The WMAP team emphasizes: ``We do not detect gravitational waves from 
inflation with 7-year WMAP data, however the upper limits are $16\%$ lower..." 
\cite{wmap7-larson}, p.11; ``The 7-year WMAP data combined with BAO 
and ${\rm H_0}$ excludes the scale-invariant spectrum by more than $3\sigma$, 
if we 
ignore tensor modes (gravitational waves)" \cite{wmap7}, p.15; ``We find no
evidence for tensor modes..." \cite{wmap7}, p.30; ``We find no
convincing deviations from the minimal model" \cite{wmap7}, p.1, etc.

In contrast, our analysis of WMAP3, WMAP5 and WMAP7 data leads us in the 
opposite direction: the improving data make the gw indications stronger. 
The major points of tension between the two approaches seem to be the constancy 
of spectral indices and the continuing use by the WMAP team of the 
inflationary theory in data analysis and interpretation. We shall start from 
the discussion of spectral indices.

The constancy of spectral indices is a reasonable assumption, but not 
a rule. If the power-law dependence (\ref{inscf}) is not a good approximation 
to the gravitational pump field during some interval of time, the constancy 
of $n_t, ~n_s$ is not a good approximation to the generated primordial spectra 
(\ref{primsp}) in the corresponding interval of wavelengths. In fact, the 
future measurements of frequency dependence of the spectrum of relic 
gravitational waves will probably be the best way to infer the ``early history 
of the Hubble parameter" \cite{grsol}.

The frequency-dependence of a gw spectrum is fully determined by the 
time-dependence of the function $\gamma(t) \equiv - {\dot H}/{H^2}$. (In more 
recent papers of other authors this function is often denoted 
$\epsilon(t)$.) The function $\gamma(t)$ describes the rate of change of the 
time-dependent Hubble radius $l_{H}(t) \equiv c/H(t)$:  
\[
\gamma(t) = \frac{d}{dt}\left(\frac{1}{H(t)}\right) = \frac{1}{c} 
\frac{dl_{H}(t)}{dt}.
\]
The function $\gamma(t)$ is a constant for power-law scale factors 
(\ref{inscf}): $\gamma = (2+\beta)/(1+\beta)$, and $\gamma=0$ for a period of 
de Sitter expansion. The interval of time $dt$ during the early era when 
gravitational waves were entering the amplifying regime and their today's 
frequency spread $d\nu$ are related by (see Eq.(21) in \cite{grsol}):
\[
\frac{d}{dt} =[1-\gamma(\nu)]H(\nu)\frac{d}{d \ln \nu}.
\]

The today's spectral energy density of gravitational waves $\epsilon(\nu)$ is 
related to the early universe parameter $\gamma(t)$ by 
(see Eq.(22) in \cite{grsol}): 
\begin{eqnarray}
\label{22}
\gamma(\nu) = -\frac{[d~\ln\epsilon(\nu)/d~\ln\nu]}
{2-[d~\ln\epsilon(\nu)/d~\ln\nu]}. 
\end{eqnarray}

The spectral index $n_g$ of a pure power-law energy density $\epsilon(\nu) 
\propto \nu^{n_g}$ is defined as $n_g = [d~\ln\epsilon(\nu)/d~\ln\nu]$.
It is reasonable to retain this definition for more complicated spectra.
Then, Eq.(\ref{22}) can be rewritten as 
\begin{eqnarray}
\label{ngnu}
n_{g}(\nu) = -\frac{2 \gamma(\nu)}{1- \gamma(\nu)}. 
\end{eqnarray}
Obviously, in the case of pure power-laws (\ref{inscf}) we return to the 
constant spectral index $n_g = -2\gamma/(1-\gamma) = 2(2+\beta)$.

Equation (\ref{ngnu}) was derived for the energy density of relatively 
high-frequency gravitational waves, $\nu > 10^{-16} {\rm Hz}$, which 
started the adiabatic regime during the radiation-dominated era. In our CMB 
study we deal with significantly lower frequencies. It is more appropriate to 
speak about wavenumbers $n$ rather than frequencies $\nu$, and about power 
spectra of metric perturbations $h^2(n)$ rather than energy density 
$\epsilon(\nu)$. The $k$-dependent spectral index $n_{t}(k)$ entering 
primordial spectrum (\ref{PsPt}) is defined as 
$n_{t}(k) = [d~\ln P_{t}(k)/d~\ln k]$. Then formula for $n_{t}(k)$ retains 
exactly the same appearance as Eq.(\ref{ngnu}):
\begin{eqnarray}
\label{ntk}
n_{t}(k) = - \frac{2 \gamma(k)}{1- \gamma(k)}. 
\end{eqnarray}
The spectral index $n_{t}(k)$ reduces to a constant $n_t=2(2+\beta)$ in the 
case of power-law functions (\ref{inscf}).

The spectral index $n_{s}(k) -1$ for density perturbations is defined by 
$n_{s}(k)-1 = [d~\ln P_{s}(k)/d~\ln k]$. The formula for $n_s(k) - 1$ 
is more complicated than Eq.(\ref{ntk}) as it contains also $d\gamma(t)/dt$ 
as a function of $k$. However, it is important to stress that adjacent 
intervals of power-law evolution (\ref{inscf}) with slightly different 
constants $\beta$, will result in slightly different pairs of constant indices
$n_t,~n_s-1$ in the corresponding adjacent intervals of wavelengths.  Of course, the spectrum itself is continues at the wavelength marking the transition between the two regions.

Extending the minimal model, the WMAP Collaboration works with the power 
spectrum \cite{wmap7} 
\begin{eqnarray}
\label{runPs}
P_{s}(k) = P_{s}(k_0) \left(\frac{k}{k_0}\right)^{n_{s}(k_0) -1 +
\frac{1}{2}\alpha_s \ln(k/k_0)},
\end{eqnarray}
which means that the $k$-dependent (running) spectral index $n_{s}(k)$ is 
assumed to be a constant plus a logarithmic correction: 
\begin{eqnarray}
\label{runns}
n_{s}(k) = n_{s}(k_0) +\alpha_s \ln(k/k_0).
\end{eqnarray}
The aim of WMAP data analysis is to find $\alpha_s$, unless it is postulated 
from the very beginning, as is done in the central (minimal) model, that 
$\alpha_s \equiv 0$. We note that although logarithmic corrections do arise 
in simple situations and can even be termed ``natural" \cite{grsol}, they are 
not unique or compulsory, as we illustrated by exact formula (\ref{ntk}). 
Nevertheless, we do not debate this point. We accept WMAP's definitions, and 
we want to illustrate their results graphically, together with our 
evaluations of $n_s$ in this paper.

\begin{figure}
{\includegraphics[height=10cm]{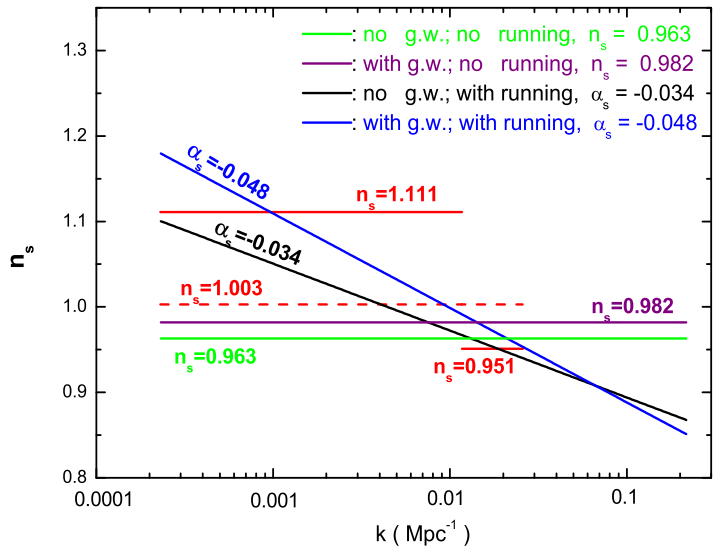}} \caption{The
spectral index $n_s$ as a function of the wavenumber $k$. The ML results 
of this work are shown by red lines. Other lines are our plots of WMAP7 
findings \cite{wmap7}, Table 7.}
\label{sec2-fig2}
\end{figure}

The main result of WMAP7 determination is $n_s=0.963 \pm 0.012$ (68\% C.L.) 
derived under the assumption of no gravitational waves and constant $n_s$
throughout all wavelengths included in the considered datasets 
\cite{wmap7,wmap7-larson}. When the presence of gravitational waves is 
allowed, but $n_s$ is still assumed constant, the $n_s$ rises to 
$n_s= 0.982^{+0.020}_{-0.019}$ from WMAP7 data alone. Finally, from WMAP7 
data alone, the WMAP team finds $n_s(k_0)=1.027,~\alpha_s= -0.034$ in the case 
of no gw but with running of $n_s$, and $n_s(k_0) =1.076,~\alpha_s = -0.048$ 
in the case of running and allowed gravitational waves (we quote only central 
values without error bars, see Table 7 in \cite{wmap7}). All the resulting 
values of $n_s(k)$ derived by WMAP team are shown in Fig.~\ref{sec2-fig2}. 
For comparison, we also plot by red lines our evaluations of $n_s$, 
see Sec.~\ref{results}.

The lines in Fig.~\ref{sec2-fig2} show clearly that our finding of 
a blue shape of the spectrum, i.e. $n_s =1.111$, at longest accessible 
wavelengths is pretty much in the territory of WMAP findings, if one 
allows running, even as simple as Eq.(\ref{runns}), and especially when 
running is combined with gravitational waves. On the other hand, as was 
already explained in \cite{stable}, the attempt of constraining relic 
gravitational waves by using the data from a huge interval of wavelengths 
and assuming a constant $n_s$ (or its simple running) across all wavelengths 
is unwarranted. The high-$\ell$ CMB data, as well as other datasets at 
relatively short wavelengths, have nothing to do with relic gravitational 
waves, and their use is dangerous. As we argued in Sec.~\ref{results}, the 
spectral index $n_s$ appears to be sufficiently different even at the span of 
two adjacent intervals of wavenumbers. The restriction to a relatively small 
number of multipoles $2 \leq \ell \leq 100$ is accompanied by relatively 
large uncertainties in $R$, but there is nothing we can do about it to 
improve the situation, this is in the nature of efforts aimed at measuring 
$R$. The difference in the treatment of $n_s$ is probably the main reason 
why we do see indications of relic gravitational waves in the data, whereas 
the WMAP team does not.

Another contributing factor to the difference of conclusions is the 
continuing use by WMAP Collaboration of the inflationary theory and its
(incorrect) relation $n_t = -r/8$, which automatically sends $r$ to zero 
when $n_t$ approaches zero. This formula is a part of the inflationary 
`consistency relations'
\[
r= 16 \epsilon = - 8 n_t.
\] 
Only one equality in this formula, $16 \epsilon = -8n_t$, is correct being
an approximate version (for small $\gamma$, $\epsilon \equiv \gamma$) of our 
exact formula, Eq. (\ref{ntk}). The `consistency relation' $r= 16 \epsilon$ is 
incorrect. It is an immediate consequence of the ``classic result" of 
inflationary theory, namely, the prediction of arbitrarily large amplitudes 
of density perturbations generated 
in the limit of de Sitter inflation ($\epsilon =0,~n_s=1$), regardless of the 
strength of the generating gravitational field (curvature of space-time) 
regulated by the Hubble parameter $H$ \footnote{It is difficult to give 
adequate references to the origin of the ``classic result". Judging from 
publications, conference talks and various interviews, there exists harsh
competition among inflationists for its authorship. One popular inflationary 
activity of present days is calculation of small loop corrections to the 
theory which is wrong by many orders of magnitude in its lowest tree 
approximation. For a more detailed criticism of inflationary theory, 
see \cite{disca}.}. Certainly, it would be inconsistent, even by the standards
of inflationary theory, not to use the relation $r= - 8n_t$ in 
data analysis, if the inflationary ``classics" is used in derivation of power 
spectra and interpretation of results (see, for example, Fig.~19 in 
\cite{wmap7}). Obviously, in our analysis we do not use the inflationary 
theory and its relations.

\section{Forecasts for the Planck mission based on the results of analysis 
of WMAP7 data\label{section4}}

The Planck satellite \cite{planck} is currently making CMB measurements and
is expected to provide data of better quality than WMAP. We hope that
the indications of relic gravitational waves that we found in WMAP3, WMAP5 and
WMAP7 data will become a certainty after Planck observations. We quantify the 
detection ability of the Planck experiment by exploring the vicinity of the WMAP7 
maximum likelihood parameters (\ref{best-fit}). 

It is seen from Fig.~\ref{figurea1.1} that the samples with relatively large 
values of the likelihood (red, yellow, and green) are concentrated 
along the curve which projects into relatively straight lines (at
least, up to $R \approx 0.5$) in the planes $R-n_s$ and $R-A_s$:
\begin{eqnarray}
\label{1Dmodel} 
n_s=0.98+0.49R, ~~~~~A_s=(2.30-1.77R)\times10^{-9}.
\end{eqnarray}
The ML model (\ref{best-fit}) is a specific point on these lines, 
corresponding to $R=0.264$. The parameterization (\ref{1Dmodel}) is close 
to the result derived from WMAP5 data: $n_s=0.98+0.46R,
~A_s=(2.27-1.53R)\times10^{-9}$, and ML $R=0.229$ (see Eq. (15) in 
\cite{stable}). We use Eq.(\ref{1Dmodel}) in formulating our forecast, thus 
reducing the task of forecasting to a 1-parameter problem in terms of $R$.

Following \cite{stable}, we define the signal-to-noise ratio as
\begin{eqnarray}
\label{snr}
 S/N\equiv R/\Delta R,
\end{eqnarray}
where the numerator is the true value of the parameter $R$ (or its
ML value, or the input value in a numerical simulation) while
$\Delta R$ in the denominator is the uncertainty in determination of $R$ 
from the data.

We estimate the uncertainty $\Delta R$ using 
the Fisher matrix formalism. We take into account all available 
information channels, i.e.  $TT$, $TE$, $EE$ and $BB$ correlation functions, 
and their various combinations. The uncertainty $\Delta R$ 
depends on instrumental and environmental noises, on the 
statistical uncertainty of the CMB signal itself, and on whether
other parameters, in addition to $R$, are derived from the same dataset. 
All our input assumptions about Planck's instrumental noises, number and 
specification of frequency channels, foreground models and residual 
contamination, sky coverage and lifetime of the mission, etc. are exactly 
the same as in our previous paper \cite{stable}. We do not repeat the 
details here and refer the reader to the text and appendices in 
\cite{stable} which contain necessary references. Technically, 
our present forecast is somewhat different (better) than that in 
the previous analysis \cite{stable} because of slightly different family 
of preferred perturbation parameters (\ref{1Dmodel}), with slightly higher 
than before the ML value of $R$, $R=0.264$. We have added only one new 
calculation, reported in Fig.~\ref{totalSNRfor28months} and 
Fig.~\ref{totalSNRforDifferentChannels2}, which is the $S/N$ for the survey of 
28 months duration, instead of the nominal assumption of 14 months.

\begin{figure}
\centerline{\includegraphics[width=18cm,height=10cm]{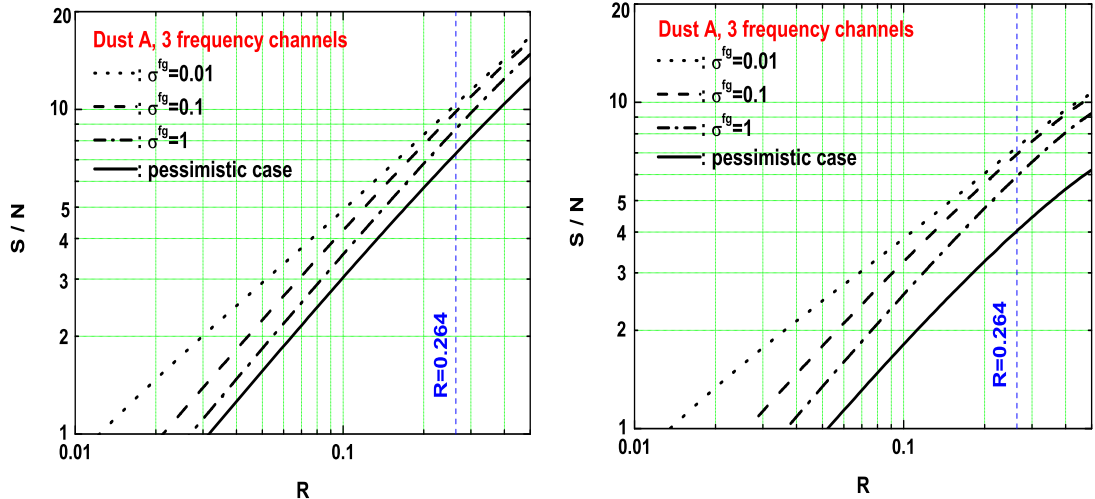}}
\caption{The total signal-to-noise ratio $S/N$ as a function of $R$.
The left panel shows $S/N$ with $\Delta R=1/\sqrt{F_{RR}}$,
while the right panel shows $S/N$ with $\Delta R=\sqrt{(F^{-1})_{RR}}$.}
\label{figurev2}
\end{figure}

The results of our forecast for the Planck mission are presented in figures. 
In Fig.~\ref{figurev2} we show the total $S/N$ for the case $TT+TE+EE+BB$, i.e. 
when all correlation functions are taken into account, and at all relevant 
multipoles $2\leq\ell\leq100$. The possible levels of foreground cleaning 
are marked by $\sigma^{\rm fg}$. The pessimistic case is the case of no 
foreground removal, $\sigma^{\rm fg}=1$, and the nominal instrumental noise
of the $BB$ information channel 
at each frequency increased by a factor of 4. Three frequency 
channels at $100$GHz, $143$GHz and $217$GHz are considered as providing 
data on perturbation parameters $R$, $n_s$, $A_s$. The more severe, Dust A, 
model is adopted for evaluation of residual foreground contamination.

In the left panel of Fig.~\ref{figurev2} we consider the idealized situation, 
where only one parameter $R$ is unknown and therefore the uncertainty 
$\Delta R$ is calculated from the $F_{RR}$ element of the Fisher matrix, 
$\Delta R=1/\sqrt{F_{RR}}$. In the right panel of Fig.~\ref{figurev2} we 
consider a more realistic situation, where all perturbation parameters 
$R$, $n_s$, $A_s$ are unknown and therefore the uncertainty $\Delta R$ 
increases and is calculated from the element of the inverse matrix,
$\Delta R=\sqrt{(F^{-1})_{RR}}$.

From the right panel of Fig.~\ref{figurev2} follows our main 
conclusion: the relic gravitational waves of the maximum likelihood  
model (\ref{best-fit}) will be detected by Planck at the healthy level 
$S/N = 4.04$, even in the pessimistic case. This is an anticipated improvement
in comparison with our evaluation $S/N = 3.65$ based on WMAP5 data analysis 
\cite{stable}. The detection will be more confident, at the level 
$S/N = 7.62,~6.91$, if $\sigma^{\rm fg}=0.01,~0.1$ can be achieved. Even in 
the pessimistic case, the signal-to-noise ratio remains at the level $S/N>2$ 
for $R>0.11$.

\begin{figure}
\begin{center}
\includegraphics[width=18cm]{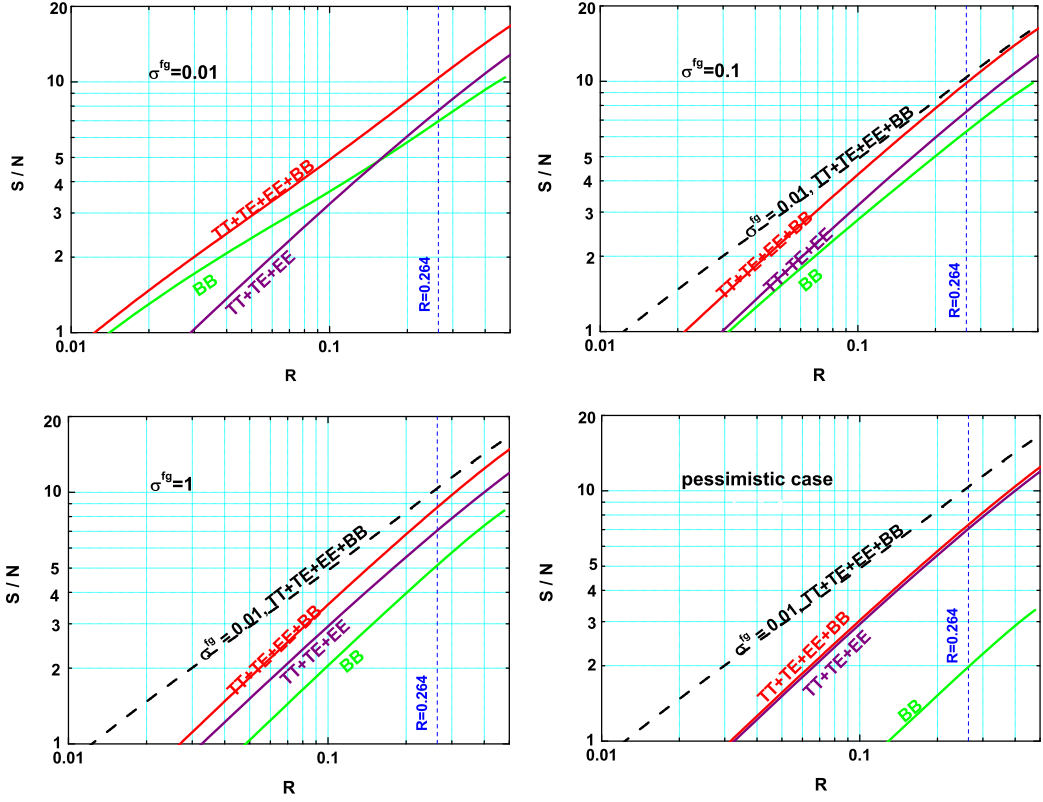}
\end{center}
\caption{The decomposition of the total $S/N$ into contributions from different 
information channels. Four panels describe different assumptions about 
foreground cleaning and noises. The $S/N$ line from the upper left panel for 
the optimistic case $\sigma^{\rm fg}=0.01$ is copied as a broken line in 
other panels.}
\label{totalSNRforDifferentChannels1}
\end{figure}

Further insight in the detection ability of Planck and 
interpretation of future results is gained by breaking up 
the total $S/N$ into contributions from different information channels 
and individual multipoles. It is easier to do this for the idealized 
situation, $\Delta R=1/\sqrt{F_{RR}}$, exhibited in the left panel of 
Fig.~\ref{figurev2}. In Fig.~\ref{totalSNRforDifferentChannels1} we
show how the $TT+TE+EE$ and $BB$ contribute to the total 
S/N based on all correlation functions $TT+TE+EE+BB$. (The $(S/N)^2$ for
the full combination $TT+TE+EE+BB$ is the sum of $(S/N)^2$ for $TT+TE+EE$
and $BB$.) It is seen from the upper left panel 
of Fig.~\ref{totalSNRforDifferentChannels1} that the 
$B$-mode of polarization is a dominant contributor to the total $S/N$ only in 
the conditions of very deep cleaning, $\sigma^{\rm fg}=0.01$, and relatively 
small values of the parameter $R$. On the other hand, in the pessimistic case, 
the $BB$ channel provides only $S/N =2.02$ for the benchmark case $R=0.264$.
Most of the total $S/N= 7.32$ in this case comes from the $TT+TE+EE$
combination.

\begin{figure}
\centerline{\includegraphics[width=18cm,height=10cm]{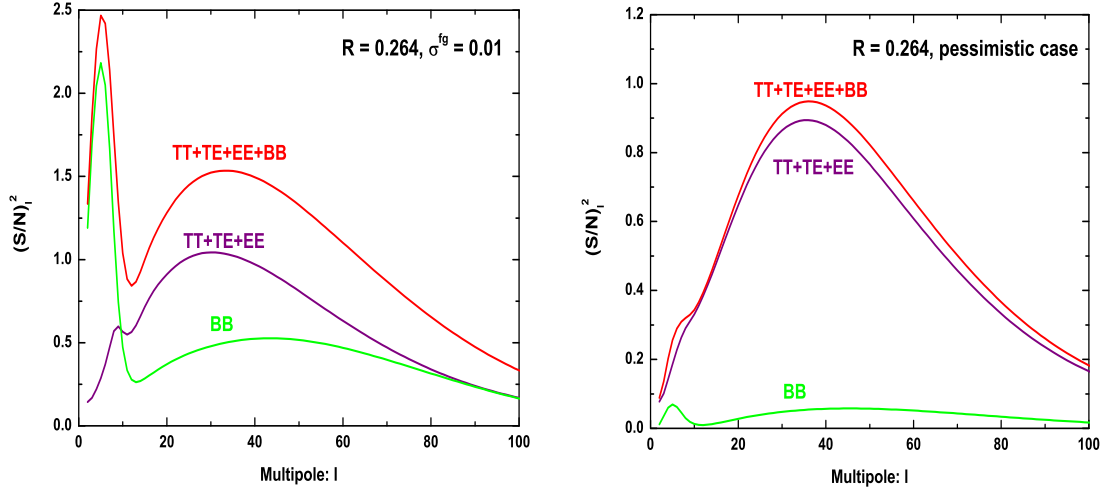}}
\caption{The individual terms $(S/N)_{\ell}^{2}$ as functions of $\ell$
for various combinations of information channels and two opposite assumptions
about residual foreground contamination and noises. The calculations are done 
for the ML model (\ref{best-fit}) with $R=0.264$.} \label{figurev3}
\end{figure}

In Fig.~\ref{figurev3} we illustrate the decomposition of the total $S/N$ into 
contributions from individual multipoles $\ell$: 
\begin{eqnarray}\label{s2}
(S/N)^2=\sum_{\ell}(S/N)_{\ell}^2.
\end{eqnarray}
We show the contributing terms $(S/N)_{\ell}^2$ for three combinations of 
information channels, $TT+TE+EE+BB$, $TT+TE+EE$ and $BB$ alone, and for two 
opposite extreme assumptions about foreground cleaning and noises, namely,  
$\sigma^{\rm fg}=0.01$ and the pessimistic case. The calculations are done
for the benchmark model (\ref{best-fit}) with $R=0.264$. Surely, the total 
$S/N$ exhibited in the upper left and lower right panels of 
Fig.~\ref{totalSNRforDifferentChannels1} is recovered with the help of
Eq.(\ref{s2}) from the sum of the terms $(S/N)_{\ell}^2$ shown in the left 
and right panels of Fig.~\ref{figurev3}, respectively.

It is seen from the left panel of Fig.~\ref{figurev3} that in the case of
deep cleaning the $BB$ channel is particularly sensitive to the very low 
multipoles $\ell\simeq 10$ associated with the reionization era. At the same 
time, the right panel of Fig.~\ref{figurev3} demonstrates that in the 
pessimistic case most of $S/N$ comes from $TT+TE+EE$ combination and, 
specifically, from the interval of mutlipoles $\ell\sim(20-60)$ associated 
with the recombination era.

\begin{figure}
\centerline{\includegraphics[width=18cm,height=10cm]{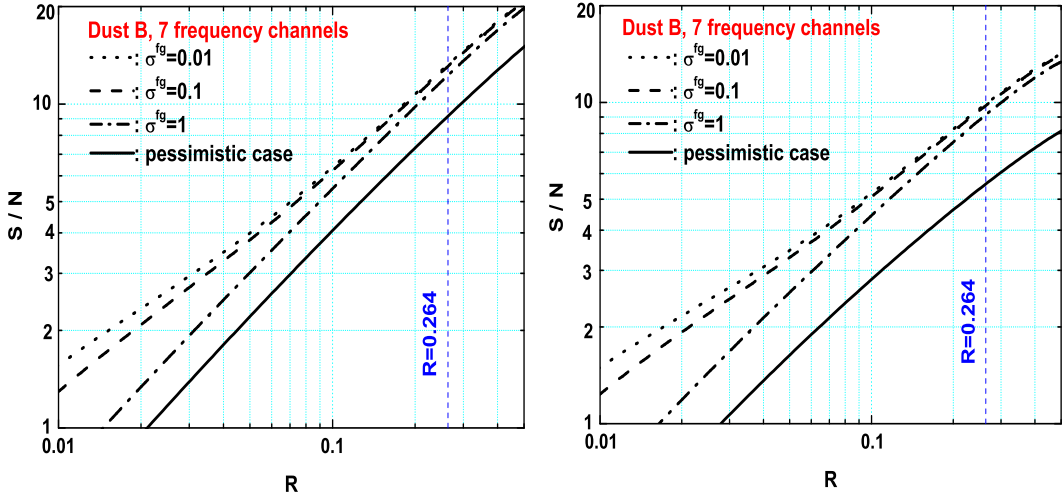}}
\caption{The improved signal-to-noise ratio $S/N$ for the scenario where 
the Dust B model is correct and the instrumental noises are smaller because
of the 7 frequency channels used, instead of 3, in data analysis. 
The left panel shows $S/N$ with $\Delta R=1/\sqrt{F_{RR}}$,
while the right panel shows $S/N$ with $\Delta R=\sqrt{(F^{-1})_{RR}}$.}
\label{figurev2b}
\end{figure}

\begin{figure}
\begin{center}
\includegraphics[width=18cm, height=10cm]{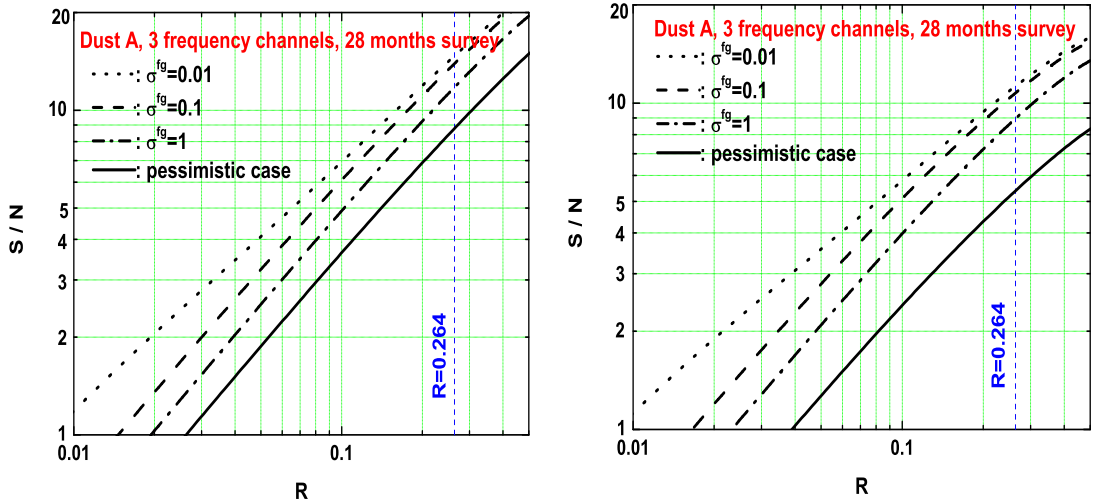}
\end{center}
\caption{The improved $S/N$ for 28 months of observations. Other assumptions
are the same as in Fig.~\ref{figurev2}.}
\label{totalSNRfor28months}
\end{figure}

\begin{figure}
\begin{center}
\includegraphics[width=12cm]{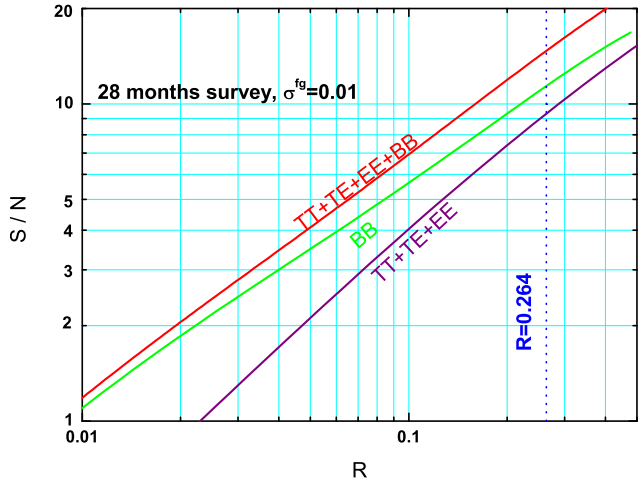}
\end{center}
\caption{Decomposition of the total $S/N$ figuring in the left panel of
Fig.~\ref{totalSNRfor28months} into contributions from different information 
channels. In the assumed conditions of deep cleaning $\sigma^{\rm fg}=0.01$, 
the $BB$ alone is more informative than the combination $TT+TE+EE$.}
\label{totalSNRforDifferentChannels2}
\end{figure}

Finally, we want to discuss possible improvements in our forecasts.
They will be achievable, if 7 frequency channels, instead of 3, 
can be used for data analysis, and/or if the less restrictive Dust B 
model, instead of the Dust A model, turns out to be correct, and/or if 28 
months of observations, instead of 14 months, can be reached. In 
Fig.~\ref{figurev2b} we show the results for the scenario where 7 frequency 
channels are used and the Dust B model is correct. Similarly to 
Fig.~\ref{figurev2}, the left panel shows $S/N$ calculated assuming that 
only $R$ is being determined from the data, whereas the right panel shows a 
more realistic case in which all perturbation parameters are unknown. In the 
later case, the $S/N$ for the ML model (\ref{best-fit}) with $R=0.264$ increases 
up to the level $S/N=5.56$ as compared with $S/N=4.04$ that we found 
in the right panel of Fig.~\ref{figurev2} for 3 frequency channels and the Dust A 
model.

The two panels in Fig.~\ref{totalSNRfor28months} illustrate the improvements,
as compared with the two panels in Fig.~\ref{figurev2}, arising from the 
longer, 28-month, survey instead of the nominal 14 months survey.
The $S/N$ for the benchmark model $R=0.264$ increases up to $S/N= 5.39$ even
in the pessimistic case, right panel. The decomposition of $S/N$ appearing in 
the left panel of Fig.~\ref{totalSNRfor28months} into contributions from 
different information channels is presented in 
Fig.~\ref{totalSNRforDifferentChannels2}. One can see that the relative role
of the $BB$ channel has increased. (The prospects of $B$-mode detection
in the conditions of low instrumental and foreground noises have been
analyzed in \cite{EG}.)

\section{Theoretical frameworks that will be accepted or rejected by Planck
observations\label{bayescomp}}

Our forecast for the Planck mission, as any forecast in nature, can prove 
its value only after actual observation. We predict sunny days 
of confident detection of relic gravitational waves, but the reality can 
turn out to be gloomy days of continuing uncertainty. One can illustrate 
this point with the help of probability distributions in
Fig.~\ref{figureb12}. We expect that Planck observations 
will continue the trend of exhibiting narrower 
likelihoods with maximum in the area near the WMAP7 ML value 
$R=0.264$. Then the detailed analysis in Sec.~\ref{section4} explains 
the reasons for our optimism. But, in principle, the reality can happen 
to be totally different. From the position of pure logic it is still 
possible that the Planck data, although making the new likelihood curve much 
narrower, will also shift the maximum of this curve to the point $R=0$. 
In this case, instead of confident detection we will have to speak about 
``tight upper limits". We do not think, though, that this is going to happen. 
What kind of conclusions about theoretical models can we make if the 
likelihood curve comes out as we anticipated?

As was already mentioned in Sec.~\ref{section2.0}, the theory of 
quantum-mechanical generation of cosmological perturbations implies a 
reasonable guess for the true value of $R$: $R\in[0.01,1]$. We shall call 
it a model $M_1$. At the same time, the most advanced string-inspired 
inflationary theories predict $R$ somewhere in the interval 
$R\in[0,10^{-24}]$. We shall call it a model $M_2$. The inflationary 
calculations can be perfectly alright in their stringy part, but the 
observational predictions are entirely hanging on the inflationary 
``classic result", and therefore should fall together with it, on purely 
theoretical grounds. But this is not the point of our present discussion. 
We wish to conduct a Bayesian comparison of models $M_1$ and $M_2$, 
regardless of the motivations that stayed behind these models.

The model $M_1$ suggests that the quadrupole ratio $R$ should lie in the range 
$[0.01,1]$ with a uniform prior in this range: $P_{\rm prior}(R|M_1)=c_1$ 
for $R\in[0.01,1]$. The model $M_2$ suggests that $R$ should lie in the range 
$[0,10^{-24}]$ with a uniform prior in this range: 
$P_{\rm prior}(R|M_2)=c_2$ for $R\in[0,10^{-24}]$. The constants $c_1$ and 
$c_2$ are determined from normalization of the prior distributions, 
$c_1= (0.99)^{-1}$ and $c_2 = 10^{24}$. The predicted interval of possible 
values of $R$ is much wider for $M_1$ than for $M_2$. Therefore, $M_1$   
is penalized by a much smaller normalization constant $c_1$ than $c_2$. 
The observed data allow one to compare the two models 
quantitatively with the help of the Bayes factor $K_{12}$ \cite{bayes}
\begin{eqnarray}
K_{12}
\equiv
\frac{
\int_0^{\infty} P_{\rm prior}(R|M_1) \mathcal{L}(R) dR
}{
\int_0^{\infty} P_{\rm prior}(R|M_2)
\mathcal{L}(R) dR},
\label{BayesFactor}
\end{eqnarray}
where $\mathcal{L}$ is the likelihood function of $R$ derived from 
the observation.

We shall start from the existing WMAP7 data. The likelihood function as a 
function of $R$ is shown by a red solid line in Fig.~\ref{figureb12}. 
Calculating the Bayes factor according to Eq.(\ref{BayesFactor}) we find
$K_{12}=1.61$, i.e. $\ln K_{12}=0.48$. As expected, this value of the Bayes 
factor, although indicative, does not provide sufficient reason for the 
rejection of model $M_2$ in favor of $M_1$.

With more accurate Planck observations the situation will change dramatically, 
if the WMAP7 maximum likelihood set of parameters (\ref{best-fit}) is correct. 
To make calculation in Eq.(\ref{BayesFactor}), we adopt a Gaussian shape of the 
likelihood function $\mathcal{L}$ with maximum at $R=0.264$ and 
$R \geq 0$. The standard deviation is taken as $\Delta R=0.065$. This value 
follows from the analysis of $S/N$ for Planck mission and corresponds to the 
derived $S/N=4.04$ in the pessimistic case. Then the calculation of $K_{12}$ 
gives the value $K_{12}=579.23$, i.e. $\ln K_{12}=6.36$. In accordance with 
Jeffrey's interpretation \cite{bayes}, this result shows that the Planck 
observations will decisively reject the model $M_2$ in favor of $M_1$.

If Planck observations are as accurate as expected, and if our assumptions 
about Planck's likelihood function are correct, the models much less 
extreme than $M_2$ will also be decisively rejected. We introduce the model 
$M_3$ with a flat prior in the range $R\in[0, x]$ and ask the question for 
which $x$ the Bayes factor $K_{13}$ exceeds the critical value $K_{13}=100$ 
($\ln K_{13}=4.61$) \cite{bayes} which is required for decisive exclusion of 
the model. The calculation gives $x=0.05$. This means that under the conditions 
listed above the Planck experiment will be able to decisively reject any 
theoretical framework that predicts $R<0.05$. In terms of the parameter $r$ 
this means the exclusion of all models with $r<0.095$.

\section{Conclusions}

The analysis of the WMAP7 data release amplifies observational indications 
in favor of relic gravitational waves in the Universe. The WMAP3, WMAP5, 
and WMAP7 temperature and polarization data in the interval of multipoles  
$2 \leq \ell \leq 100$ persistently point out to one and the same area 
in the space of perturbation parameters. It includes a considerable amount 
of gravitational waves expressed in terms of the parameter $R=0.264$, and 
somewhat blue primordial spectra with indices $n_s = 1.111$ and 
$n_t =0.111$. If the maximum likelihood set of parameters that we derived 
from this analysis is a fair representation of the reality, the relic 
gravitational waves will be detected more confidently by Planck observations. 
Even under pessimistic assumption about hindering factors, the expected 
signal-to-noise ratio should be at the level $S/N = 4.04$.

~

{\bf Acknowledgements}
We acknowledge the use of the LAMBDA and CAMB Websites.
W. Z. is partially supported by Chinese NSF Grants No.
10703005, and No. 10775119 and the Foundation for University
Excellent Young Teacher by the Ministry of Zhejiang Education.


\end{document}